\documentclass[11pt,preprint,superscriptaddress,nofootinbib]{revtex4}
\usepackage{amsmath,amssymb,graphicx,booktabs,tabularx,siunitx}
\usepackage{url}
\begin{document}

\title{Trilepton Selection and Kinematic Sensitivity for Multiple Weak-Boson Production at Hadron Colliders}
\author{Ijaz Ahmed}
\email{Ijaz.Ahmed@fuuast.edu.pk}
\affiliation{Federal Urdu University of Arts, Science and Technology, Islamabad, Pakistan}
\author{Jamil Muhammad}
\affiliation{Sang-Ho College of Education, Konkuk University, Seoul 05029, Korea}
\email{jamil.konkuk@gmail.com}
\author{M. S. Amjad}
\affiliation{National University of Technology, Islamabad, Pakistan}
\author{M. Tayyab Javaid}
\affiliation{Federal Urdu University of Arts, Science and Technology, Islamabad, Pakistan}
\author{Farzana Ahmad}
\affiliation{SETI, College of Engineering, Konkuk University, Seoul 05029, South Korea}

\date{\today}

\begin{abstract}
We present a detector-level fast-simulation study of the charge-specific processes $pp\to W^+W^-W^+$ and $pp\to W^+W^-W^-$ in the trilepton channel at 8, 14, 27, and 100 TeV. Separate signal regions require exactly three isolated electrons or muons with net charge $+1$ or $-1$, respectively, together with ordered lepton transverse-momentum thresholds, a same-flavour opposite-sign $Z$ veto, missing transverse momentum, and restricted jet multiplicity. For the $W^+W^-W^+$ hypothesis, the expected signal/background yields at reference luminosities of 20 fb$^{-1}$, 3 ab$^{-1}$, 15 ab$^{-1}$, and 20 ab$^{-1}$ are $(1.27,0.186)$, $(180,40.7)$, $(2.81\times10^3,784)$, and $(1.09\times10^4,3.79\times10^3)$, with conditional Asimov significances of 1.84, 7.70, 8.62, and 6.31. For $W^+W^-W^-$, the corresponding signal efficiencies range from $9.0\times10^{-4}$ to $2.95\times10^{-3}$ and the nominal significances are 1.19, 8.02, 8.13, and 6.03. The inclusive charge-channel asymmetry decreases from 0.351 at 8 TeV to 0.166 at 100 TeV; its quoted uncertainty is limited to numerical cross-section integration errors. Exploratory likelihood, neural-network, and boosted-tree classifiers show no stable improvement over rectangular cuts. The projections are conditional on the supplied simulated samples and uncertainty model; important backgrounds and comprehensive detector and theory uncertainties are not included.
\end{abstract}

\maketitle

\section{Introduction}
Multiple weak-boson production provides a direct way to test the structure of electroweak interactions. The non-Abelian $SU(2)_L\times U(1)_Y$ gauge symmetry fixes the Standard Model interactions among the gauge bosons, including the charged $WW\gamma$ and $WWZ$ triple-gauge vertices (TGVs) and the quartic gauge vertices. Precision measurements of these interactions test the gauge structure of the theory and the cancellations among amplitudes that control high-energy scattering. Departures from the Standard Model predictions could reveal indirect effects of new degrees of freedom, commonly parameterized through anomalous couplings or higher-dimensional operators in an effective field theory (EFT) description \cite{HagiwaraTGC,SMEFTbasis}.

Diboson production, particularly $WW$, $WZ$, and $W\gamma$, provides the conventional direct sensitivity to charged TGVs through the relevant electroweak production amplitudes and differential distributions. Triboson production extends this program: its amplitudes contain contributions from electroweak gauge self-interactions and can probe quartic gauge vertices as well as provide complementary information on triple-gauge interactions. The $WWW$ final state is especially useful because its charge and lepton-multiplicity patterns provide a distinctive experimental signature; it has been observed in proton--proton collisions by ATLAS \cite{ATLASWWW}. The rarity of these processes makes them challenging, but also motivates their study at higher collision energies, where larger event samples and harder kinematic spectra can improve access to the electroweak interaction structure. Their interpretation nevertheless requires differential predictions, adequate background control, and an EFT validity treatment when non-Standard-Model operators are considered \cite{KubotaTGCQGC,TripleWQGC}.

We study $pp\to W^+W^-W^+$ (SM-T1) and $pp\to W^+W^-W^-$ (SM-T2) at centre-of-mass energies of 8, 14, 27, and 100 TeV. Our aim is to establish Standard Model cross-section and trilepton sensitivity benchmarks and compare cut-based with exploratory multivariate selections. Coupling limits are outside the scope of this study.

\section{Samples, simulation, and normalization}
\label{sec:samples}
The process set contains the supplied diboson samples, same-sign $WWjj$ samples generated through electroweak and QCD contributions, triboson samples, and four-boson samples. We perform two separate signal hypotheses: SM-T1 in the net-charge $+1$ region and SM-T2 in the net-charge $-1$ region. For each hypothesis, the opposite-charge triboson sample and all other readable processes are backgrounds. For the SM-T1 result, process-group labels are Diboson, Same-sign $WWjj$, Other triboson, Four-boson, and the charge-conjugate SM-T2 contribution.

For each sample, ROOT event weights are constructed from the corresponding integrated cross section in MG5\_aMC@NLO production cross-section records \cite{MG5}, the number of readable generated events, and the assumed integrated luminosity. For a sample with cross section $\sigma_i$, $N_i^{\rm gen}$ generated entries, and $n_i$ entries passing a selection, the predicted yield is
\begin{equation}
  N_i = \mathcal{L}\,\sigma_i\,\frac{n_i}{N_i^{\rm gen}}.
  \label{eq:yield}
\end{equation}
The cross-section integration uncertainty is propagated as a sample-wise normalization uncertainty. Yields are normalized to readable ROOT entries: all 15 process trees are readable at each energy, and every tree contains 100,000 entries except SM-Q3 at 100 TeV, which has 40,496. The reference luminosities are 20, 3,000, 15,000, and 20,000 fb$^{-1}$ at 8, 14, 27, and 100 TeV, respectively, and are listed with the signal results.

The Delphes files were scanned for reconstructed electrons, muons, jets, and missing transverse momentum \cite{Delphes}. Candidate leptons were formed using $p_T\geq10$ GeV, $|\eta|<2.5$, and the Delphes branch requirement \texttt{IsolationVar}<0.15. Jets were counted for $p_T>30$ GeV and $|\eta|<4.5$. These are fast-simulation object selections; a dedicated trigger, pile-up overlay, detector calibration, and fake-lepton model are not included.

\section{Event selection and optimization}
\label{sec:selection}
The SM-T1 baseline region requires exactly three selected electrons or muons with net reconstructed charge $+1$. The SM-T2 analysis uses the same baseline with net charge $-1$. The leptons are ordered by transverse momentum and must satisfy $p_T^{\ell_1}\geq25$ GeV, $p_T^{\ell_2}\geq15$ GeV, and $p_T^{\ell_3}\geq10$ GeV. To reduce events containing a resonant $Z$ boson, we reject an event if any same-flavour opposite-sign (SFOS) pair has $|m_{\ell\ell}-m_Z|<\Delta_Z$. We then impose a missing transverse momentum threshold, an $H_T$ threshold, and an upper bound on reconstructed jet multiplicity. Here $H_T$ is the scalar sum of selected jet transverse momenta.

The scan uses discrete thresholds: $\Delta_Z=0,10,15,20,25,30$ GeV; $E_T^{\rm miss}=0,20,30,40,50,60,80,100,120$ GeV; $H_T=0,50,100,150,200,300,500$ GeV; and $N_j^{\max}=10,4,3,2,1,0$. Even-numbered event entries are used to choose the cuts and odd-numbered entries provide an independent validation. Candidate cuts are required to retain minimum signal and background counts and minimum effective background simulation statistics in both folds. The selected point maximizes the smaller of the two fold significances.

\begin{table}[ht]
\centering\small
\caption{Optimized trilepton selections and projected yields. $\epsilon_S$ is the fraction of generated SM-T1 events passing all stated cuts. $Z_A$ is the Asimov profile-likelihood significance including the assumed 10\% background normalization uncertainty, finite-MC uncertainty, and MG5 integration errors.}
\label{tab:optimized}
\resizebox{\textwidth}{!}{%
\begin{tabular}{|r|r|r|r|r|r|r|r|r|r|r|}\hline
$\sqrt{s}$ & $\mathcal{L}$ & $\Delta_Z$ & $E_T^{\rm miss}$ & $H_T$ & $N_j^{\max}$ & $\epsilon_S$ & $S$ & $B$ & $S/B$ & $Z_A$ \\ \hline
(TeV) & (fb$^{-1}$) & (GeV) & (GeV) & (GeV) & & & & & & \\ \hline
8 & 20 & 10 & 60 & 0 & 2 & $1.86\times10^{-3}$ & 1.27 & 0.186 & 6.80 & 1.84 \\ \hline
14 & 3000 & 10 & 80 & 0 & 1 & $7.30\times10^{-4}$ & 180 & 40.7 & 4.42 & 7.70 \\ \hline
27 & 15000 & 30 & 50 & 0 & 1 & $9.40\times10^{-4}$ & $2.81\times10^3$ & 784 & 3.59 & 8.62 \\ \hline
100 & 20000 & 20 & 20 & 0 & 2 & $5.90\times10^{-4}$ & $1.09\times10^4$ & $3.79\times10^3$ & 2.89 & 6.31 \\ \hline
\end{tabular}%
}
\end{table}

\begin{table}[ht]
\centering\small
\caption{Cumulative cut flow at $\sqrt{s}=14$ TeV and $\mathcal{L}=3000$ fb$^{-1}$. Signal efficiency is relative to all generated SM-T1 events; yields use the MG5 cross sections and readable ROOT entry counts. The final row applies the optimized missing-momentum and jet requirements.}
\label{tab:cutflow}
\begin{tabular}{|l|r|r|r|}\hline
Cumulative requirement & $\epsilon_S$ & $S$ events & $B$ events \\ \hline
$\geq3$ isolated $e/\mu$ ($p_T\geq10$ GeV, $|\eta|<2.5$) & $\num{0.00305}$ & $\num{751}$ & $\num{7.63e+04}$ \\ \hline
Exactly three selected leptons & $\num{0.00304}$ & $\num{749}$ & $\num{5.54e+04}$ \\ \hline
Net charge $+1$ & $\num{0.00304}$ & $\num{749}$ & $\num{2.88e+04}$ \\ \hline
Ordered lepton thresholds $25,15,10$ GeV & $\num{0.003}$ & $\num{739}$ & $\num{2.85e+04}$ \\ \hline
SFOS $Z$ veto ($\Delta_Z=10$ GeV) & $\num{0.00255}$ & $\num{628}$ & $\num{4.89e+03}$ \\ \hline
Optimized: $E_T^{\rm miss}\geq80$ GeV, $H_T\geq0$ GeV, $N_j\leq1$ & $\num{0.00073}$ & $\num{180}$ & $\num{40.7}$ \\ \hline
\end{tabular}
\end{table}

The selected $H_T$ threshold is zero at every energy: within this scan, imposing a positive $H_T$ threshold did not improve the robust significance. The optimized jet requirements are at most one or two selected jets, depending on energy.

\section{Kinematic distributions and cut performance}
\label{sec:results}
\begin{figure}[p]
\centering\includegraphics[width=0.97\textwidth]{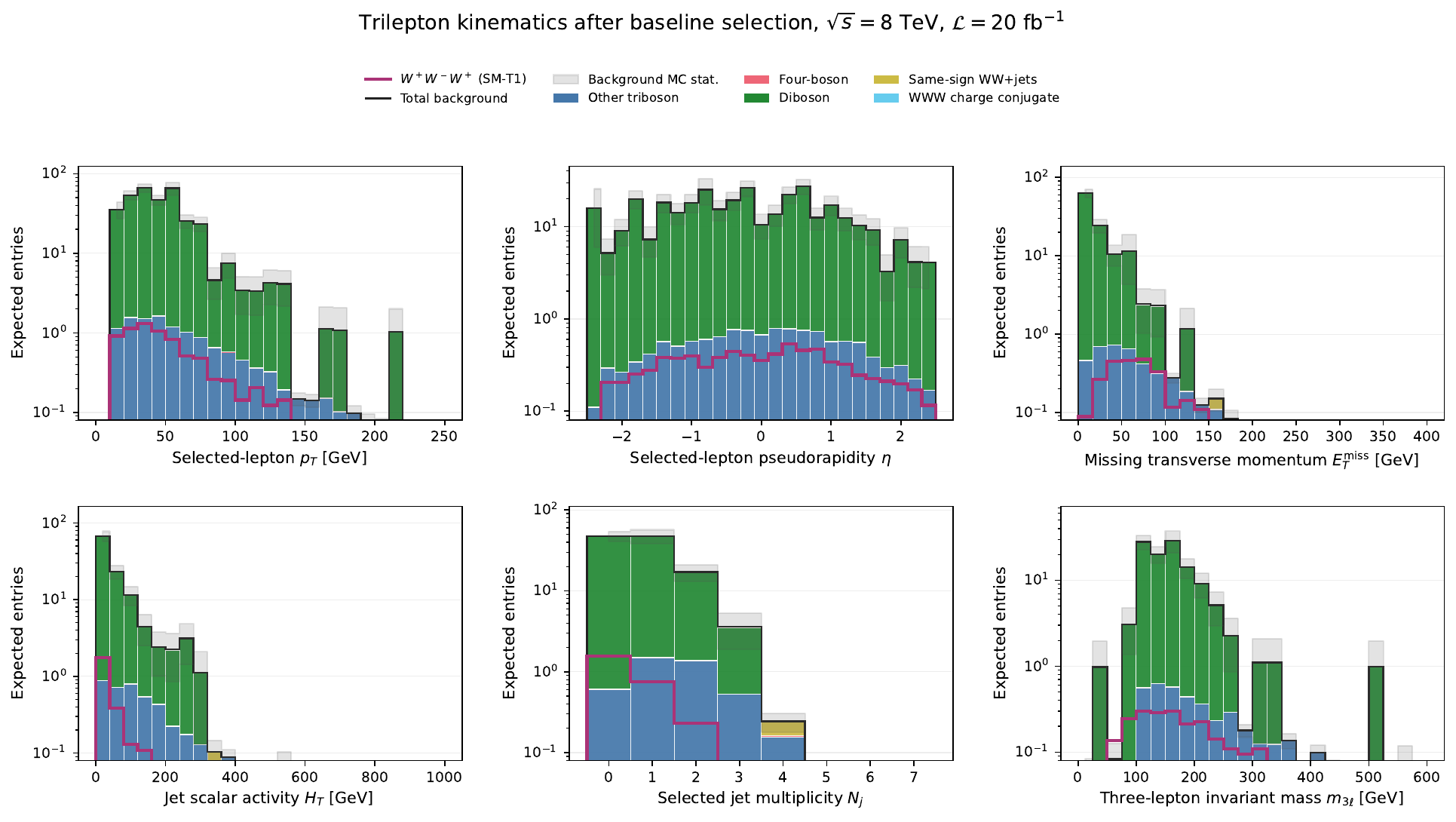}
\caption{Selected-lepton and event-level kinematics at $\sqrt{s}=8$ TeV for $20$ fb$^{-1}$. The baseline requires exactly three isolated leptons, net charge $+1$, and ordered thresholds $p_T^{\ell_1,\ell_2,\ell_3}\geq25,15,10$ GeV. The SFOS veto and optimized $E_T^{\rm miss}$ and jet cuts have not yet been applied.}
\label{fig:kin8}
\end{figure}

\begin{figure}[p]
\centering\includegraphics[width=0.97\textwidth]{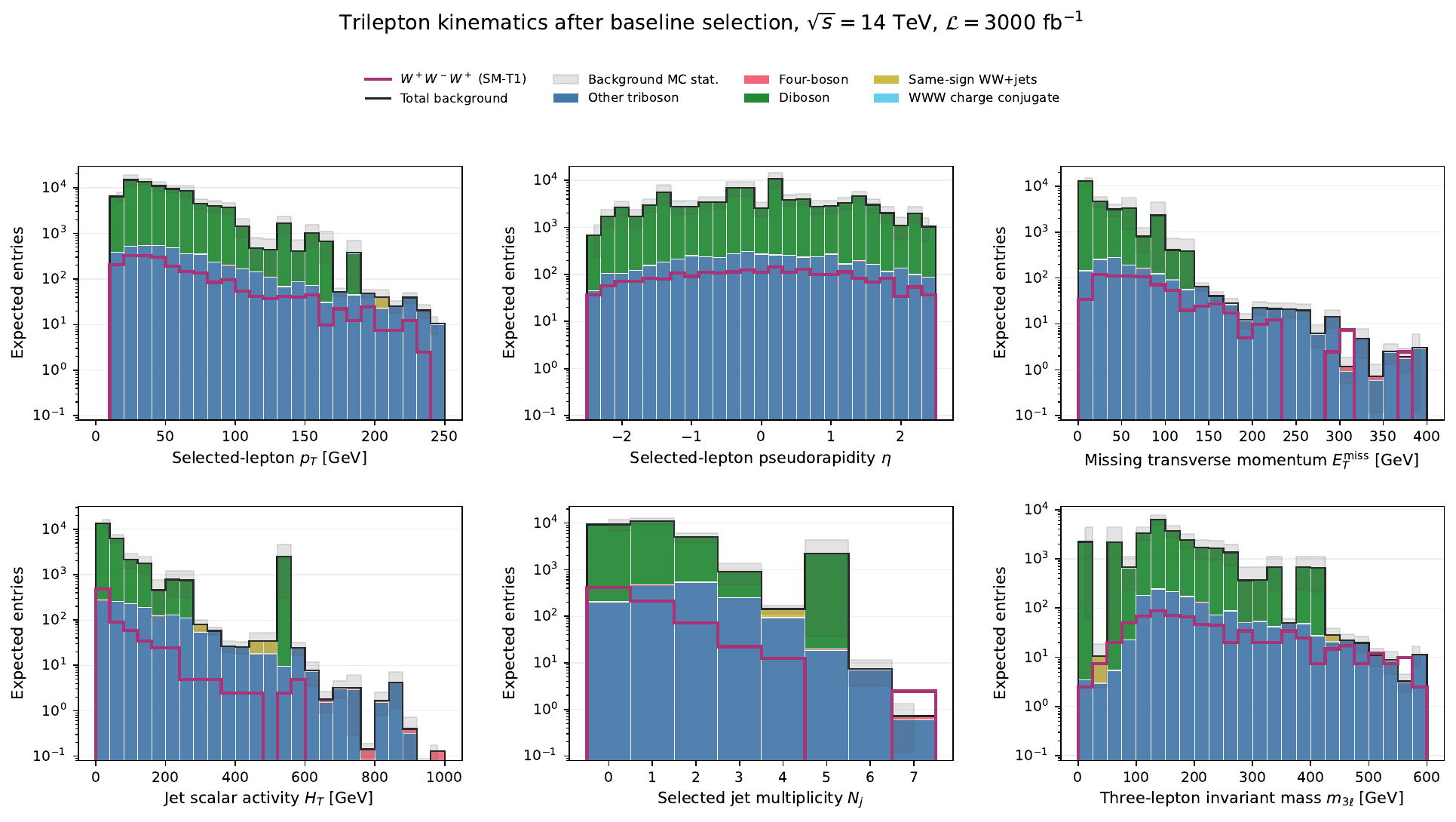}
\caption{Selected-lepton and event-level kinematics at $\sqrt{s}=14$ TeV for $3$ ab$^{-1}$, with the same preselection and plotting conventions as Fig.~\ref{fig:kin8}.}
\label{fig:kin14}
\end{figure}

\begin{figure}[p]
\centering\includegraphics[width=0.97\textwidth]{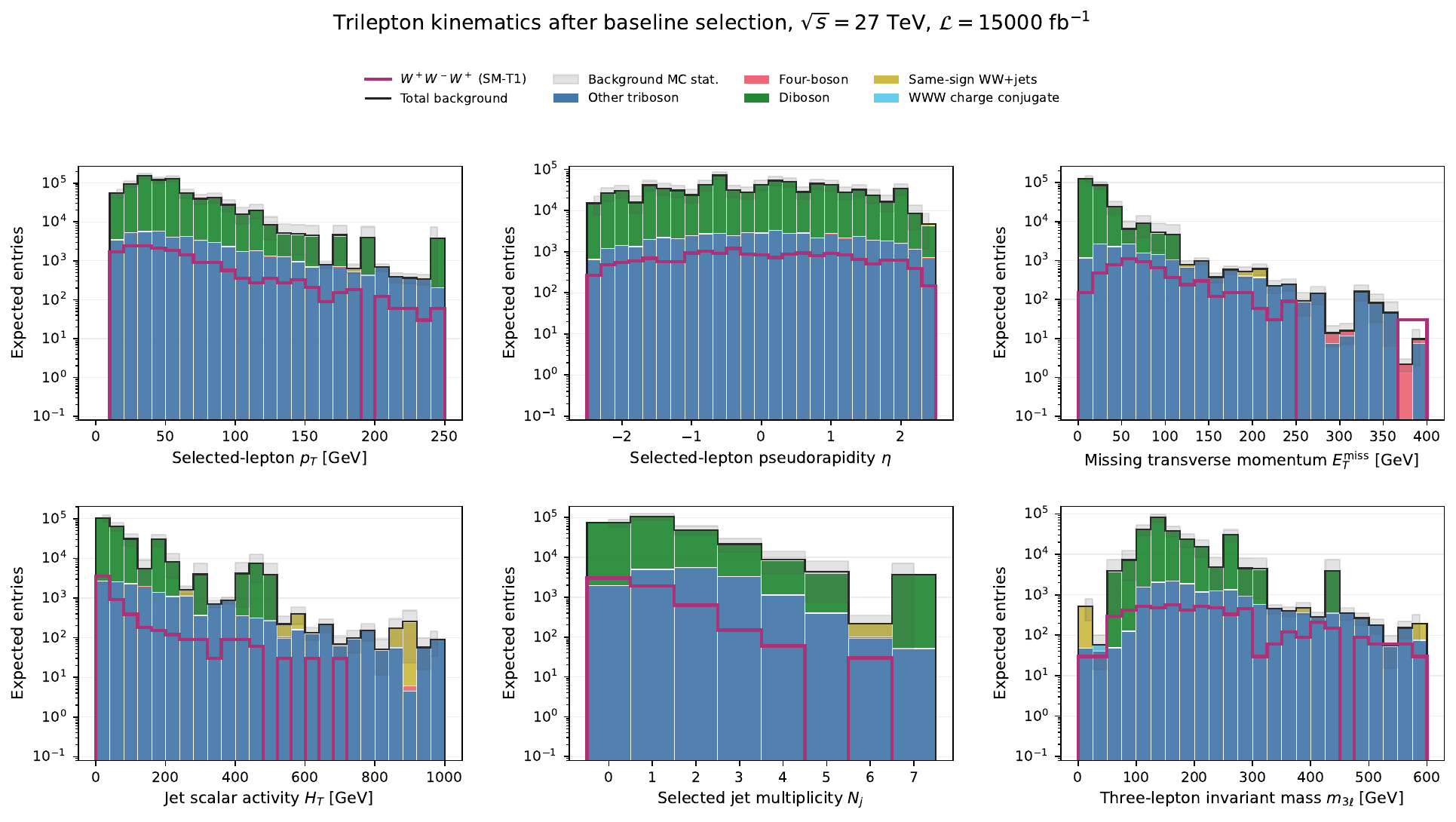}
\caption{Selected-lepton and event-level kinematics at $\sqrt{s}=27$ TeV for $15$ ab$^{-1}$, using the same net-charge $+1$ preselection and plotting conventions as Fig.~\ref{fig:kin8}.}
\label{fig:kin27}
\end{figure}

\begin{figure}[p]
\centering\includegraphics[width=0.97\textwidth]{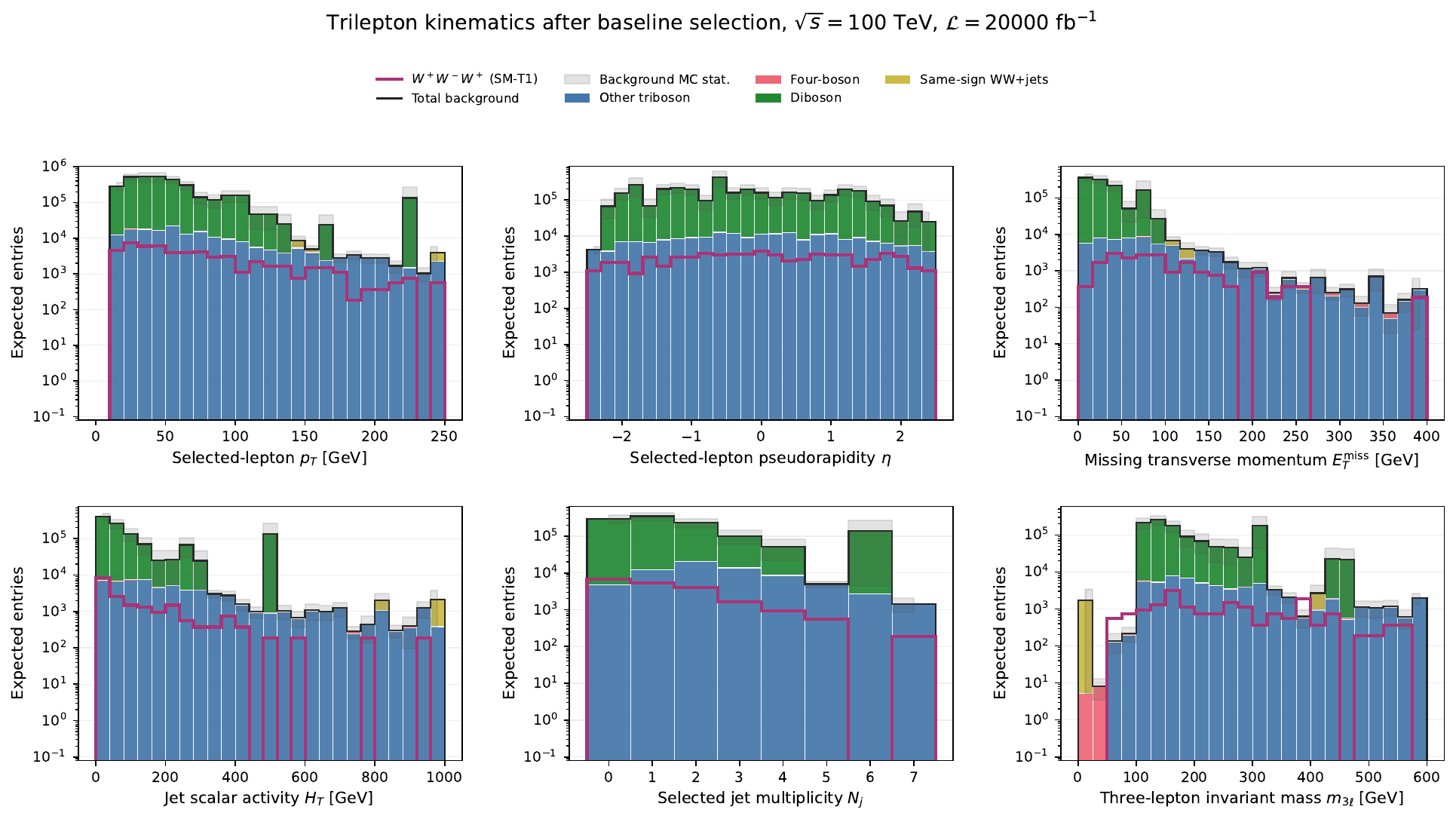}
\caption{Selected-lepton and event-level kinematics at $\sqrt{s}=100$ TeV for $20$ ab$^{-1}$, using the same net-charge $+1$ preselection and plotting conventions as Fig.~\ref{fig:kin8}.}
\label{fig:kin100}
\end{figure}

Figure~\ref{fig:kin8}--Fig.~\ref{fig:kin100} show the expected distributions after the three-lepton, charge, and ordered-$p_T$ baseline, before the SFOS veto and optimized missing-momentum and jet requirements. The six focused panels show the selected-lepton $p_T$ and pseudorapidity, $E_T^{\rm miss}$, jet $H_T$, jet multiplicity, and three-lepton invariant mass. Lepton-level entries include all three selected leptons per event; event-level variables count each event once. The stacked backgrounds are grouped by process class, the SM-T1 signal is overlaid, and the shaded band indicates background MC statistical uncertainty.

The lepton $p_T$ and $\eta$ spectra characterize the fiducial acceptance and show substantial signal--background overlap. Missing transverse momentum, which reflects the neutrinos from the $W$ decays, is a useful optimization variable. Jet multiplicity and $H_T$ describe accompanying hadronic activity, while the three-lepton invariant mass is shown as a diagnostic rather than used as a mass-window cut.


\begin{figure}[ht]
\centering\includegraphics[width=0.76\textwidth]{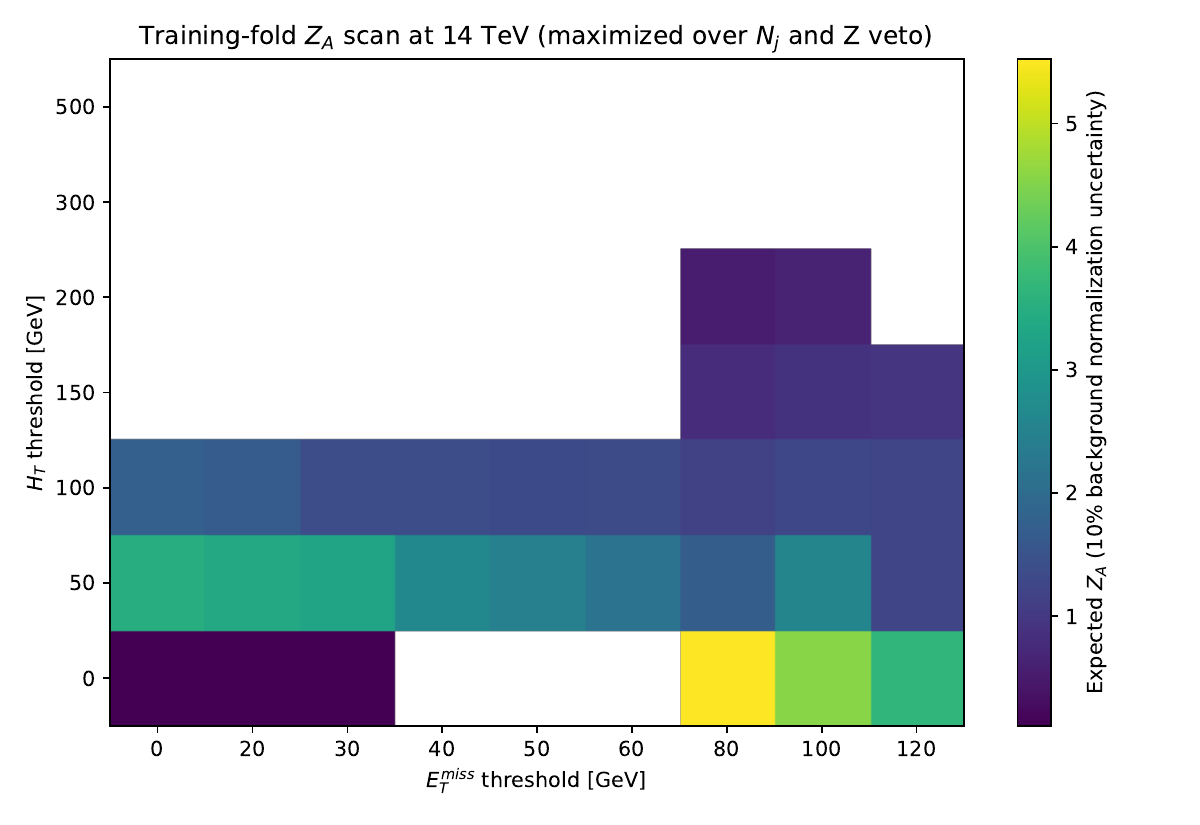}
\caption{Training-fold significance scan at 14 TeV. Each cell shows the best training-fold $Z_A$ over the scanned $Z$-veto and maximum-jet thresholds for a given pair of $E_T^{\rm miss}$ and $H_T$ thresholds. The final selection is the robust point selected using both folds.}
\label{fig:scan14}
\end{figure}

The two-dimensional cut-scan maps in Fig.~\ref{fig:scan14} illustrate the training-fold significance over missing transverse momentum and $H_T$, with the jet and $Z$-veto thresholds optimized at each point. The selected point was checked on the independent fold. At 8 TeV, the training and validation fold significances are 1.82 and 1.83; at 14 TeV they are 5.52 and 6.23; at 27 TeV they are 6.25 and 6.60; at 100 TeV they are 4.97 and 4.25. This validation reduces, but does not eliminate, the statistical variability associated with finite event samples.

\begin{table}[ht]
\centering\small
\caption{Process-group contributions to the selected yields for the SM-T1 hypothesis. A zero-valued cell means that no simulated event from that group passed the selection; it does not imply a zero physical background.}
\label{tab:groups}
\begin{tabular}{|r|r|r|r|r|r|r|}\hline
$\sqrt{s}$ (TeV) & SM-T1 signal & Other triboson & Four-boson & SM-T2 & Diboson & Same-sign $WWjj$ \\ \hline
8 & 1.269 & 0.175 & 0.0111 & 0 & 0 & 0 \\ \hline
14 & 179.858 & 38.836 & 1.868 & 0 & 0 & 0 \\ \hline
27 & 2814.078 & 743.011 & 22.551 & 18.203 & 0 & 0 \\ \hline
100 & 10932.228 & 3460.612 & 326.679 & 0 & 0 & 0 \\ \hline
\end{tabular}
\end{table}

The selected background is dominated by other triboson samples, especially $W^+W^-Z$, with a smaller four-boson contribution. Empty bins in some samples reflect limited simulated statistics; the finite-MC uncertainty is included in $Z_A$.

\begin{figure}[ht]
\centering\includegraphics[width=0.76\textwidth]{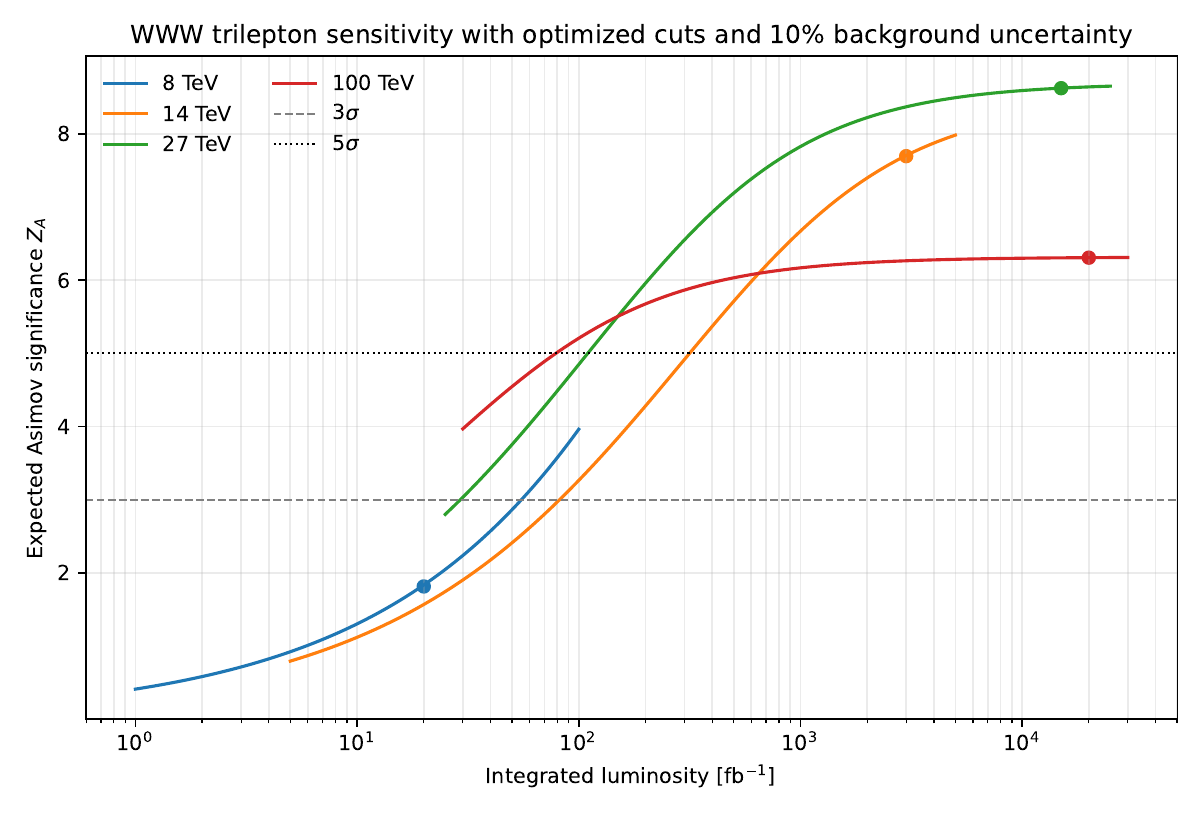}
\caption{Projected Asimov significance versus integrated luminosity at each centre-of-mass energy, using the corresponding optimized selection and a 10\% background normalization uncertainty. Curves rescale the available MC efficiencies and cross sections; they do not account for energy-dependent detector conditions or additional backgrounds.}
\label{fig:lumi}
\end{figure}

\section{Cross sections and representative production topologies}
\begin{figure}[ht]
\centering\includegraphics[width=0.92\textwidth]{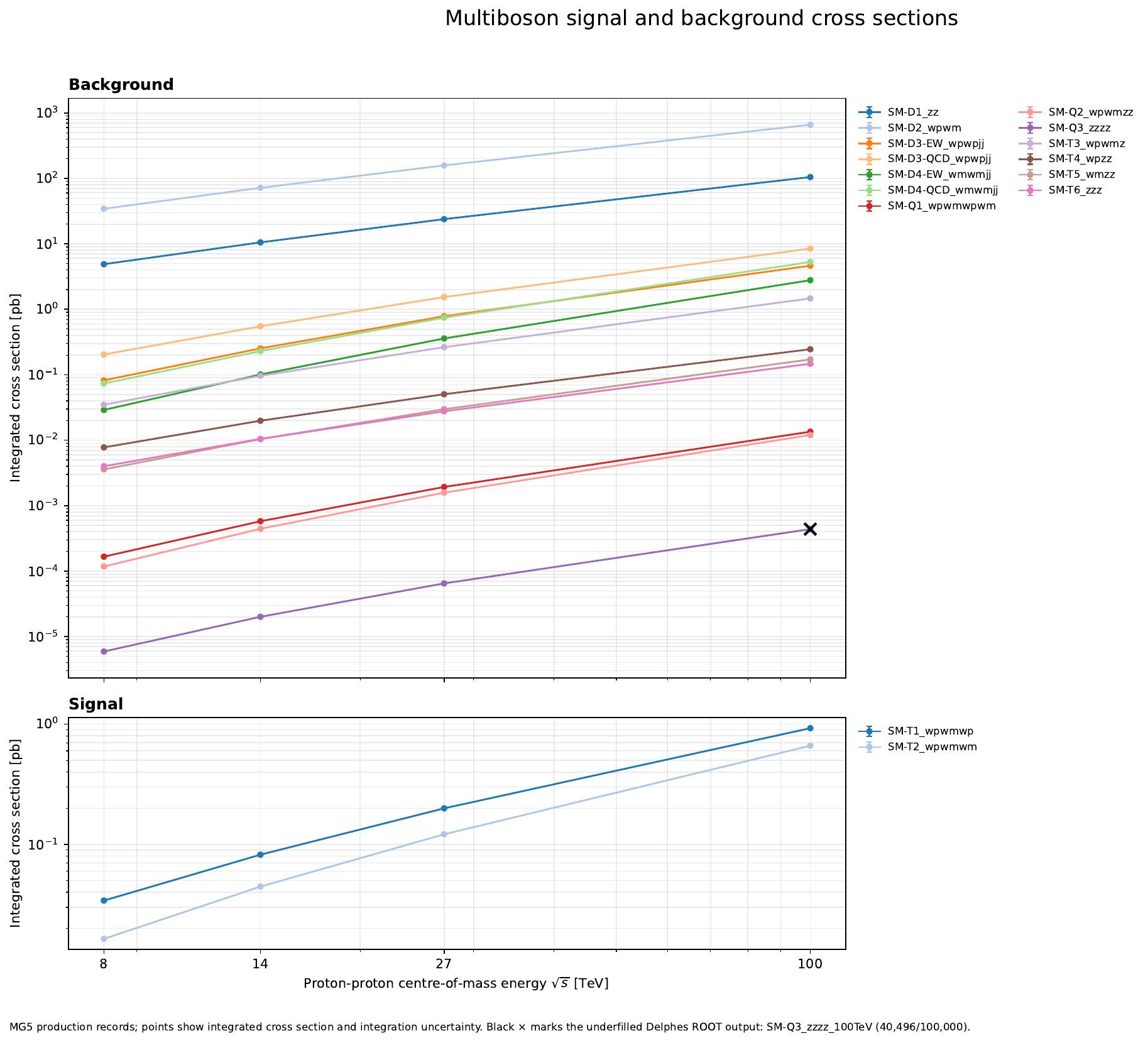}
\caption{Summed generated cross sections versus centre-of-mass energy for the supplied process groups, constructed from the MG5 production records used to normalize the ROOT samples. The charge-specific SM-T1 and SM-T2 signals are shown separately. These inclusive rates are distinct from the selected yields, which also include decay, reconstruction, and event-selection efficiencies.}
\label{fig:xsec}
\end{figure}

The complete MG5 cross-section grid, including integration uncertainties, is reported in Tables~\ref{tab:xsec:diboson}--\ref{tab:xsec:four-boson}; Fig.~\ref{fig:xsec} summarizes its energy dependence. The inclusive rates differ from the selected trilepton yields in Tables~\ref{tab:optimized} and~\ref{tab:t2results}, which also reflect decay, reconstruction, and selection efficiencies.

\begin{table*}[ht]
\centering\scriptsize
\setlength{\tabcolsep}{4pt}
\caption{MG5 integrated production cross sections for the Diboson and same-sign $WWjj$ sample grid. Entries are $\sigma\pm\delta\sigma$ in pb; $\delta\sigma$ is the reported integration uncertainty.}
\resizebox{\textwidth}{!}{%
\begin{tabular}{|l|c|c|c|c|}\hline
Process & $8$ TeV & $14$ TeV & $27$ TeV & $100$ TeV \\ \hline
$pp\to ZZ$ (SM-D1) & $\num{4.8973}\pm\num{0.0017894}$ & $\num{10.543}\pm\num{0.0036599}$ & $\num{23.869}\pm\num{0.0089921}$ & $\num{105.01}\pm\num{0.029841}$ \\ \hline
$pp\to W^+W^-$ (SM-D2) & $\num{34.394}\pm\num{0.009273}$ & $\num{71.921}\pm\num{0.027504}$ & $\num{158.08}\pm\num{0.049224}$ & $\num{662.93}\pm\num{0.18735}$ \\ \hline
$pp\to W^+W^+jj$ (EW) (SM-D3-EW) & $\num{0.082207}\pm\num{7.4759e-05}$ & $\num{0.25295}\pm\num{0.00025243}$ & $\num{0.78485}\pm\num{0.000656}$ & $\num{4.6414}\pm\num{0.0035673}$ \\ \hline
$pp\to W^+W^+jj$ (QCD) (SM-D3-QCD) & $\num{0.20432}\pm\num{0.00019915}$ & $\num{0.55096}\pm\num{0.00050851}$ & $\num{1.5378}\pm\num{0.001323}$ & $\num{8.4633}\pm\num{0.0062974}$ \\ \hline
$pp\to W^-W^-jj$ (EW) (SM-D4-EW) & $\num{0.029032}\pm\num{3.0671e-05}$ & $\num{0.101}\pm\num{0.00010807}$ & $\num{0.35878}\pm\num{0.00034295}$ & $\num{2.7884}\pm\num{0.002313}$ \\ \hline
$pp\to W^-W^-jj$ (QCD) (SM-D4-QCD) & $\num{0.073507}\pm\num{7.1906e-05}$ & $\num{0.2303}\pm\num{0.00021972}$ & $\num{0.74567}\pm\num{0.00067697}$ & $\num{5.2951}\pm\num{0.0046246}$ \\ \hline
\end{tabular}%
}
\label{tab:xsec:diboson}
\end{table*}
\begin{table*}[ht]
\centering\scriptsize
\setlength{\tabcolsep}{4pt}
\caption{MG5 integrated production cross sections for the Triboson sample grid. Entries are $\sigma\pm\delta\sigma$ in pb; $\delta\sigma$ is the reported integration uncertainty.}
\resizebox{\textwidth}{!}{%
\begin{tabular}{|l|c|c|c|c|}\hline
Process & $8$ TeV & $14$ TeV & $27$ TeV & $100$ TeV \\ \hline
$pp\to W^+W^-W^+$ (SM-T1, signal) & $\num{0.0341}\pm\num{3.0404e-05}$ & $\num{0.082127}\pm\num{8.3434e-05}$ & $\num{0.19958}\pm\num{0.00017143}$ & $\num{0.92646}\pm\num{0.00077464}$ \\ \hline
$pp\to W^+W^-W^-$ (SM-T2) & $\num{0.01638}\pm\num{1.3636e-05}$ & $\num{0.044575}\pm\num{4.4874e-05}$ & $\num{0.12135}\pm\num{0.00010217}$ & $\num{0.66224}\pm\num{0.00058614}$ \\ \hline
$pp\to W^+W^-Z$ (SM-T3) & $\num{0.03491}\pm\num{3.3143e-05}$ & $\num{0.096446}\pm\num{8.6891e-05}$ & $\num{0.26397}\pm\num{0.00022812}$ & $\num{1.4659}\pm\num{0.0011346}$ \\ \hline
$pp\to W^+ZZ$ (SM-T4) & $\num{0.0077785}\pm\num{8.5447e-06}$ & $\num{0.019854}\pm\num{2.0877e-05}$ & $\num{0.050443}\pm\num{5.0026e-05}$ & $\num{0.24545}\pm\num{0.00017043}$ \\ \hline
$pp\to W^-ZZ$ (SM-T5) & $\num{0.0035845}\pm\num{3.5445e-06}$ & $\num{0.0104}\pm\num{9.758e-06}$ & $\num{0.029798}\pm\num{3.132e-05}$ & $\num{0.17182}\pm\num{0.00014991}$ \\ \hline
$pp\to ZZZ$ (SM-T6) & $\num{0.0040047}\pm\num{2.8951e-06}$ & $\num{0.010478}\pm\num{8.5516e-06}$ & $\num{0.027673}\pm\num{2.524e-05}$ & $\num{0.14727}\pm\num{0.00012367}$ \\ \hline
\end{tabular}%
}
\label{tab:xsec:triboson}
\end{table*}
\begin{table*}[ht]
\centering\scriptsize
\setlength{\tabcolsep}{4pt}
\caption{MG5 integrated production cross sections for the Four-boson sample grid. Entries are $\sigma\pm\delta\sigma$ in pb; $\delta\sigma$ is the reported integration uncertainty.}
\resizebox{\textwidth}{!}{%
\begin{tabular}{|l|c|c|c|c|}\hline
Process & $8$ TeV & $14$ TeV & $27$ TeV & $100$ TeV \\ \hline
$pp\to W^+W^-W^+W^-$ (SM-Q1) & $\num{0.00016624}\pm\num{1.7159e-07}$ & $\num{0.00057874}\pm\num{5.7204e-07}$ & $\num{0.0019307}\pm\num{1.8587e-06}$ & $\num{0.013487}\pm\num{1.1436e-05}$ \\ \hline
$pp\to W^+W^-ZZ$ (SM-Q2) & $\num{0.00011757}\pm\num{1.3001e-07}$ & $\num{0.00044356}\pm\num{4.8218e-07}$ & $\num{0.0015814}\pm\num{1.498e-06}$ & $\num{0.01201}\pm\num{1.0541e-05}$ \\ \hline
$pp\to ZZZZ$ (SM-Q3) & $\num{5.9186e-06}\pm\num{6.8032e-09}$ & $\num{2.0053e-05}\pm\num{1.9236e-08}$ & $\num{6.4944e-05}\pm\num{6.4004e-08}$ & $\num{0.0004381}\pm\num{4.3764e-07}$ \\ \hline
\end{tabular}%
}
\label{tab:xsec:four-boson}
\end{table*}

The uncertainties in Tables~\ref{tab:xsec:diboson}--\ref{tab:xsec:four-boson} are the numerical integration errors reported by MG5. They do not include renormalization- or factorization-scale variation, parton-distribution-function uncertainty, or higher-order corrections. The values therefore document the generated leading-order sample normalization and should not be read as full theory uncertainties.

\section{Charge-channel cross-section asymmetry}
\label{sec:chargeasym}
The separate SM-T1 and SM-T2 production samples allow us to quantify the inclusive charge-channel asymmetry before detector selection. We define the cross-section ratio and normalized asymmetry as
\begin{equation}
R_{WWW}(\sqrt{s}) = \frac{\sigma_{T1}}{\sigma_{T2}}, \qquad
A_C^{WWW}(\sqrt{s}) = \frac{\sigma_{T1}-\sigma_{T2}}{\sigma_{T1}+\sigma_{T2}}
= \frac{R_{WWW}-1}{R_{WWW}+1},
\label{eq:chargeasym}
\end{equation}
where $\sigma_{T1}=\sigma(pp\to W^+W^-W^+)$ and $\sigma_{T2}=\sigma(pp\to W^+W^-W^-)$. These are inclusive production cross sections from the generated samples; the quantities in this section do not include trilepton branching fractions, detector acceptance, or optimized-selection efficiencies.

\begin{figure}[ht]
\centering\includegraphics[width=0.94\textwidth]{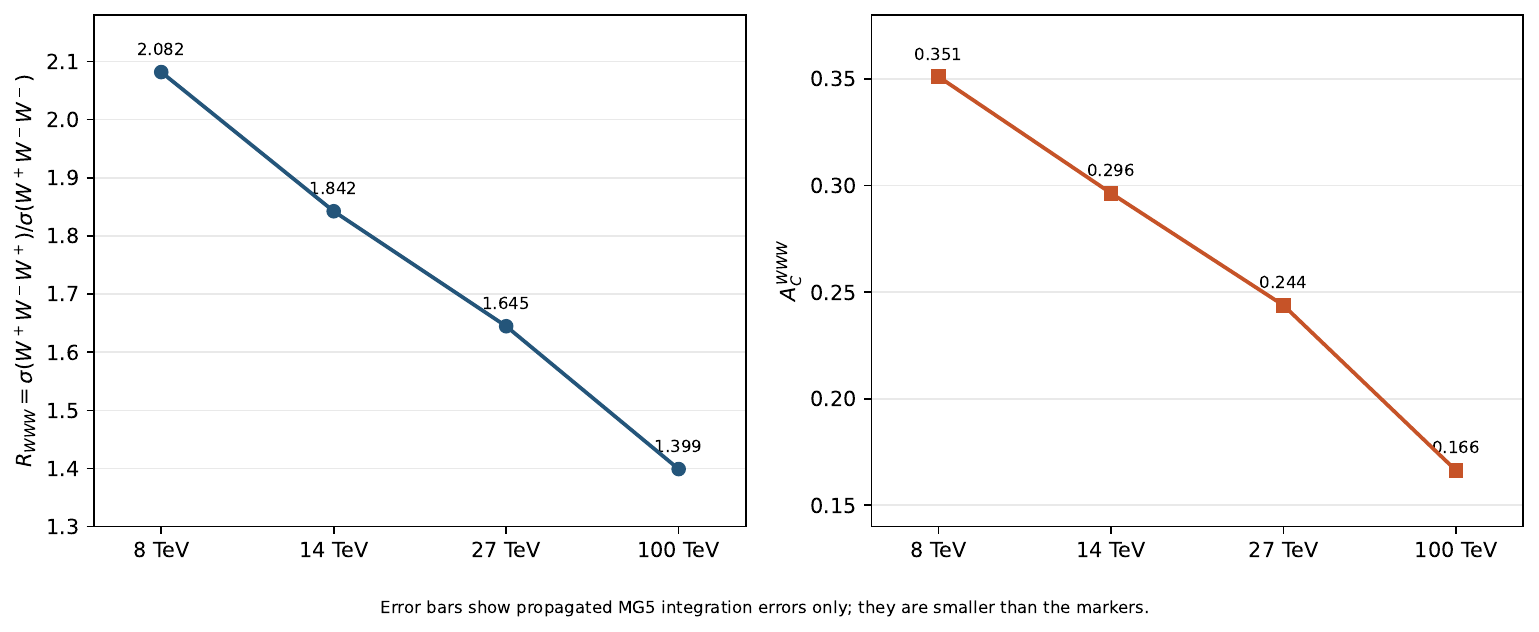}
\caption{Inclusive SM-T1/SM-T2 cross-section ratio (left) and normalized charge-channel asymmetry (right) versus proton--proton centre-of-mass energy. Error bars propagate MG5 integration errors only and are smaller than the markers; PDF, scale, and higher-order uncertainties are not included.}
\label{fig:chargeasym}
\end{figure}

The uncertainties in Table~\ref{tab:chargeasym} propagate the quoted MG5 integration errors, treating the two cross-section integrations as statistically independent. For $R_{WWW}$ and $A_C^{WWW}$, respectively, we use
\begin{align}
\delta R_{WWW} &= R_{WWW}\sqrt{\left(\frac{\delta\sigma_{T1}}{\sigma_{T1}}\right)^2+\left(\frac{\delta\sigma_{T2}}{\sigma_{T2}}\right)^2},\\
\delta A_C^{WWW} &= \frac{2\sqrt{\sigma_{T2}^{2}(\delta\sigma_{T1})^{2}+\sigma_{T1}^{2}(\delta\sigma_{T2})^{2}}}{(\sigma_{T1}+\sigma_{T2})^{2}}.
\end{align}
Both $R_{WWW}$ and $A_C^{WWW}$ decrease monotonically over the studied energy range, as shown in Fig.~\ref{fig:chargeasym} and quantified in Table~\ref{tab:chargeasym}.

This positive asymmetry is expected in proton--proton collisions because the proton's quark and antiquark flavor distributions are not charge-conjugation symmetric. Its energy dependence is therefore sensitive in principle to the flavor-weighted parton luminosities entering the two charge channels. Charge-asymmetry measurements in inclusive $W$ production are used to test proton PDFs \cite{ATLASWCharge}; the triple-$W$ observable is a complementary, higher-multiplicity prediction. The decreasing asymmetry in our samples is qualitatively consistent with a reduced valence-quark imbalance at the smaller momentum fractions increasingly sampled at higher collision energies, but this interpretation cannot be established from the four inclusive rates alone. In particular, $A_C^{WWW}$ is a production charge asymmetry, not a CP-violating observable: the $pp$ initial state is not a CP eigenstate.

As discussed for the inclusive cross-section grid above, these propagated errors include only numerical integration; PDF and scale variations, higher-order corrections, and experimental effects remain unaccounted for. The table and plot provide a useful Standard Model benchmark from the existing samples; a precision study would require matched higher-order predictions, PDF error sets, and a fiducial unfolding or detector-level likelihood. This observable also does not by itself constrain a triple-gauge coupling: that requires coupling-varied predictions and a dedicated fit.

\begin{table}[ht]
\centering\small
\caption{Inclusive charge-channel cross sections and derived asymmetry for the two triboson hypotheses. Uncertainties on the cross sections are the quoted MG5 integration errors; uncertainties on $R_{WWW}$ and $A_C^{WWW}$ propagate those errors, assuming independent integrations. They do not include PDF, scale, or other theory uncertainties.}
\label{tab:chargeasym}
\begin{tabular}{|c|c|c|c|c|}\hline
$\sqrt{s}$ (TeV) & $\sigma_{T1}$ (pb) & $\sigma_{T2}$ (pb) & $R_{WWW}=\sigma_{T1}/\sigma_{T2}$ & $A_C^{WWW}$ \\ \hline
8 & $0.0341\pm3.04\times10^{-5}$ & $0.01638\pm1.36\times10^{-5}$ & $2.082\pm0.003$ & $0.3510\pm0.0005$ \\ \hline
14 & $0.082127\pm8.34\times10^{-5}$ & $0.044575\pm4.49\times10^{-5}$ & $1.842\pm0.003$ & $0.2964\pm0.0007$ \\ \hline
27 & $0.19958\pm1.71\times10^{-4}$ & $0.12135\pm1.02\times10^{-4}$ & $1.645\pm0.002$ & $0.2438\pm0.0006$ \\ \hline
100 & $0.92646\pm7.75\times10^{-4}$ & $0.66224\pm5.86\times10^{-4}$ & $1.399\pm0.002$ & $0.1663\pm0.0006$ \\ \hline
\end{tabular}
\end{table}

\begin{figure}[ht]
\centering\includegraphics[width=0.82\textwidth]{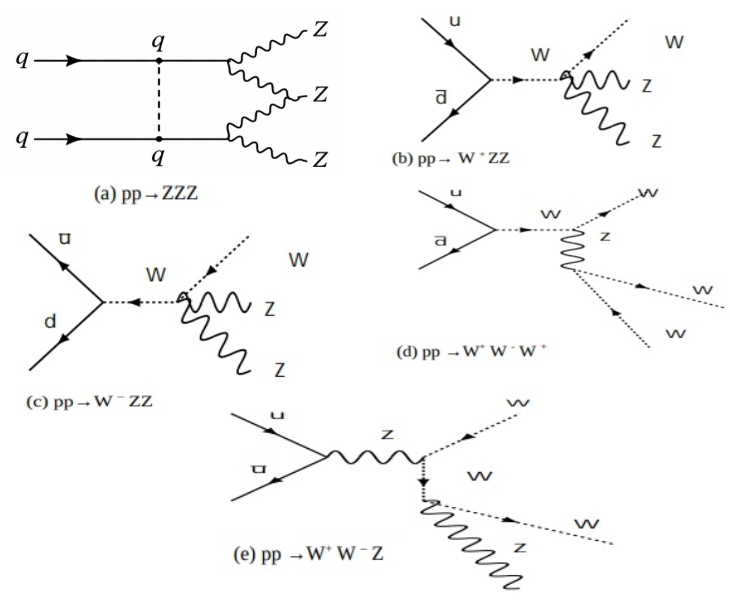}
\caption{Representative Feynman-diagram topologies for multiple weak-boson production, reproduced from the pre-existing manuscript. The analysis treats the two $WWW$ charge channels as separate Standard Model signal hypotheses.}
\label{fig:diagrams}
\end{figure}

Figure~\ref{fig:diagrams} shows representative tree-level production topologies for multiple weak bosons. The two signal hypotheses studied here are the Standard Model charge channels $W^+W^-W^+$ and $W^+W^-W^-$.

\section{Exploratory ROOT-based multivariate analysis}
\label{sec:mva}

To assess whether multivariate classifiers can improve the trilepton discrimination, we performed an exploratory study with the ROOT Toolkit for Multivariate Analysis (TMVA) \cite{TMVA}. The SM-T1 input samples first pass the reconstructed-object baseline: exactly three isolated electrons or muons, net charge $+1$, and ordered lepton thresholds of 25, 15, and 10 GeV. The classifier inputs are the transverse momenta and pseudorapidities of the three selected leptons, missing transverse momentum, jet $H_T$, jet and $b$-jet multiplicities, three-lepton invariant mass, the distance of the closest SFOS pair from the $Z$ mass, an SFOS-pair indicator, and the minimum lepton--missing-momentum azimuthal separation. The SFOS distance is capped at 300 GeV to limit extreme values.

Three TMVA methods are compared: the one-dimensional probability-density estimator (Likelihood), a multilayer perceptron (MLP), and a boosted decision tree (BDT). At each energy, event entries are assigned deterministically to training (60\%), operating-point calibration (20\%), and final test (20\%) folds. The classifier threshold is selected using only the calibration fold; the reported AUC and yields are evaluated on the held-out test fold. The rectangular-cut benchmark is optimized on the same calibration fold and evaluated on the same test fold. Training weights preserve relative cross sections among samples within each class, while the total signal and background class weights are equalized for training. Expected yields are restored to the original cross-section normalization and scaled to the reference luminosities listed with the SM-T1 results in Table~\ref{tab:optimized}.

\begin{table}[ht]
\centering\scriptsize
\caption{Exploratory ROOT TMVA comparison on the held-out test fold. $Z_A$ includes the finite-simulation and MG5 integration uncertainties and the assumed 10\% background normalization uncertainty. $B_{\rm eff}$ is the effective number of background MC events contributing after the classifier or cut threshold. The final column gives actual passing test-fold MC counts ($n_S/n_B$), illustrating the limited tail statistics.}
\label{tab:mva}
\resizebox{\textwidth}{!}{%
\begin{tabular}{|c|l|r|r|r|r|r|c|}\hline
$\sqrt{s}$ & Method & $Z_A$ & Test AUC & $S$ & $B$ & $B_{\rm eff}$ & Test MC ($n_S/n_B$) \\ \hline
(TeV) & & & & & & & \\ \hline
8 & Rectangular cuts & 1.623 & -- & 0.955 & 0.114 & 3.51 & 28/14 \\ \hline
8 & Likelihood & 0.127 & 0.756 & 0.716 & 5.05 & 1.06 & 21/46 \\ \hline
8 & MLP & 0.108 & 0.708 & 0.853 & 10.6 & 2.34 & 25/127 \\ \hline
8 & BDT & 0.255 & 0.964 & 2.08 & 10.5 & 2.28 & 61/130 \\ \hline
14 & Rectangular cuts & 1.170 & -- & 24.6 & 15.4 & 1.13 & 2/12 \\ \hline
14 & Likelihood & 0.098 & 0.560 & 160 & $1.66\times10^3$ & 1.10 & 13/50 \\ \hline
14 & MLP & 5.012 & 0.570 & 185 & 22.6 & 2.28 & 15/30 \\ \hline
14 & BDT & 0.047 & 0.760 & 530 & $1.41\times10^4$ & 1.64 & 43/53 \\ \hline
27 & Rectangular cuts & 4.917 & -- & $4.04\times10^3$ & $1.07\times10^3$ & 5.83 & 27/26 \\ \hline
27 & Likelihood & 0.066 & 0.453 & $4.79\times10^3$ & $2.67\times10^5$ & 15.91 & 32/496 \\ \hline
27 & MLP & 0.049 & 0.611 & 898 & $1.87\times10^4$ & 1.09 & 6/50 \\ \hline
27 & BDT & 3.377 & 0.981 & $4.04\times10^3$ & $2.22\times10^3$ & 7.88 & 27/84 \\ \hline
100 & Rectangular cuts & 1.717 & -- & $3.71\times10^3$ & $2.42\times10^3$ & 2.50 & 4/20 \\ \hline
100 & Likelihood & 0.056 & 0.500 & $1.85\times10^4$ & $1.02\times10^6$ & 10.52 & 20/294 \\ \hline
100 & MLP & 0.953 & 0.695 & $1.85\times10^3$ & $1.57\times10^3$ & 1.15 & 2/9 \\ \hline
100 & BDT & 2.420 & 0.966 & $1.02\times10^4$ & $6.89\times10^3$ & 5.41 & 11/29 \\ \hline
\end{tabular}%
}
\end{table}

\begin{figure}[ht]
\centering\includegraphics[width=0.88\textwidth]{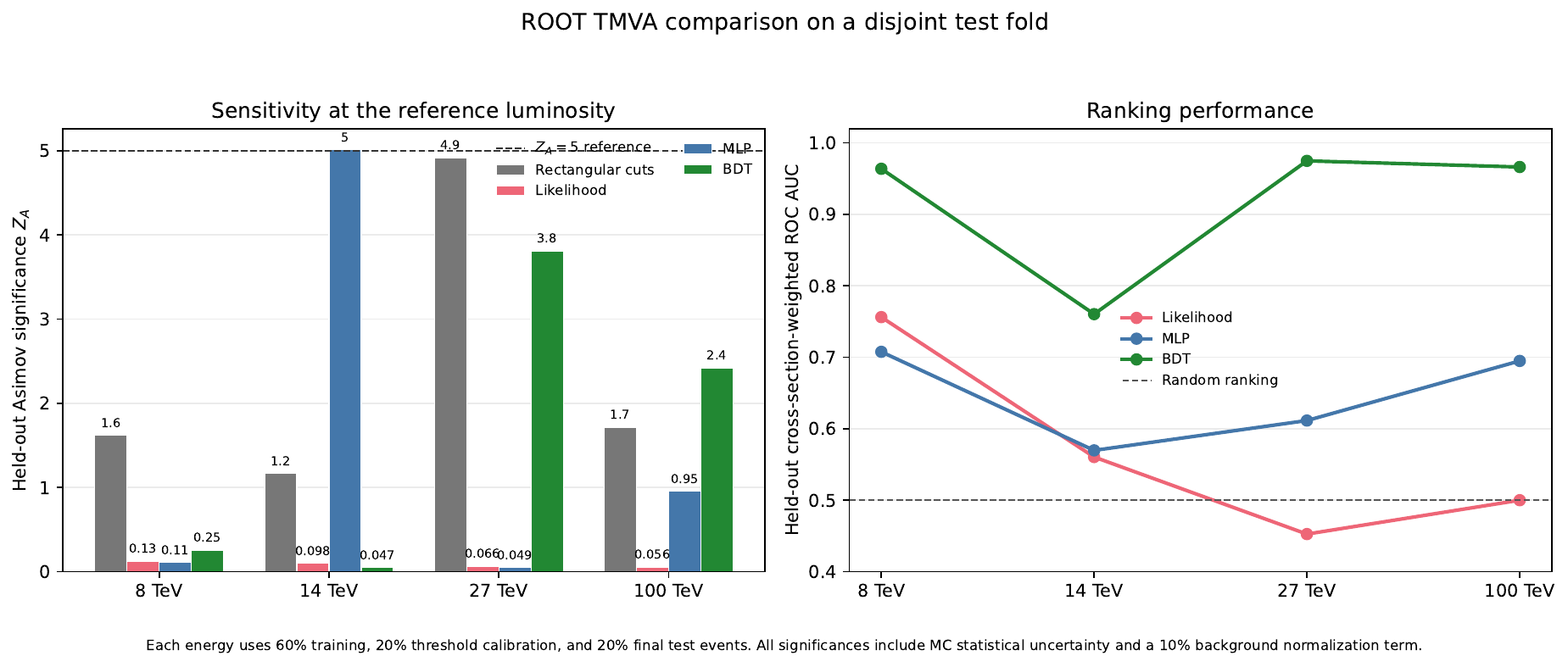}
\caption{Held-out ROOT TMVA comparison at the reference luminosities. Left: Asimov significance for the calibration-selected operating point, evaluated on the independent test fold. Right: cross-section-weighted test-fold ROC AUC; the dashed line marks random ranking. The significance panel includes a $Z_A=5$ guide, not a discovery claim. Small and energy-dependent test samples, especially in the selected tails, make these points statistically unstable.}
\label{fig:mva}
\end{figure}

The comparison does not establish a robust multivariate gain over rectangular cuts because the apparent gains are not consistent across energies and are supported by sparse test-fold tails. At 8 TeV, all three MVA test-fold significances (0.11--0.26) are below the rectangular-cut value (1.62). At 14 TeV, the MLP gives a larger nominal value ($Z_A=5.01$ versus 1.17 for cuts), but its test AUC is only 0.570 and the selected test fold contains 15 signal and 30 background events. At 27 TeV, the BDT value ($Z_A=3.38$) is below the cut benchmark (4.92), with 27 signal and 84 background test events passing, and $B_{\rm eff}=7.88$. At 100 TeV, the BDT is higher than the cut benchmark (2.42 versus 1.72), but the estimate is based on 11 signal and 29 background test events and an effective background count of 5.41. These small counts mean that a few events or their weights can materially change the selected yields and significance. The BDT training AUC is close to unity at every energy, while its held-out AUC falls to 0.760 at 14 TeV, which is evidence of overtraining. AUC and significance probe different aspects of performance: a strong global ranking does not guarantee a stable high-purity operating point when the simulated tail is sparsely populated.

The test-fold yields are rescaled by the inverse test-fold fraction to estimate full-sample yields, but the selected test events remain a finite, statistically fluctuating subset. This explains why the test-fold rectangular-cut significances in Table~\ref{tab:mva} need not equal the primary full-sample significances in Table~\ref{tab:optimized}; the former are an independent-fold diagnostic, while the latter are the central projections from the cut-optimization workflow. Larger independent samples, repeated stratified cross-validation, and a frozen operating point tested on an independent production are needed to determine whether any MVA gain is reproducible. The available production and processing resources did not permit generating the additional independent event samples required for that study. We therefore report Table~\ref{tab:mva} and Fig.~\ref{fig:mva} as an exploratory method comparison, not as the primary sensitivity result or a replacement for the cut-based projections. A realistic multivariate sensitivity study must also include major omitted backgrounds and detector-related uncertainties discussed in Sec.~\ref{sec:scope}.

\section{Charge-conjugate SM-T2 signal results}
\label{sec:t2}
We analyze $pp\to W^+W^-W^-$ (SM-T2) as a second, independent signal hypothesis. Its trilepton region requires exactly three isolated electrons or muons with reconstructed net charge $-1$ and the same ordered lepton thresholds of 25, 15, and 10 GeV. The SM-T1 process and all other supplied samples form the background in this region. This charge-specific treatment avoids mixing the two hypotheses and applies the same event-level methodology to each.

Table~\ref{tab:t2results} summarizes the cross sections, optimized selections, efficiencies, yields, and significance estimates for SM-T2. Its nominal significance exceeds 5 at 14, 27, and 100 TeV, while the independent validation values are lower, most notably at 100 TeV. This variation underscores the limited stability of projections based on the available samples.

\begin{table}[ht]
\centering\scriptsize
\caption{Charge-conjugate SM-T2 trilepton results, treating $pp\to W^+W^-W^-$ as the signal and requiring net charge $-1$. The efficiency is relative to the 100,000 generated entries. $S$ and $B$ are projected at the stated luminosity; $\delta B$ combines finite-MC, cross-section integration, and the assumed 10\% background normalization uncertainties. The scan uses the even/odd event split for cut selection and validation.}
\label{tab:t2results}
\resizebox{\textwidth}{!}{%
\begin{tabular}{|r|r|l|r|r|r|r|r|r|r|}\hline
$\sqrt{s}$ & $\mathcal{L}$ & Selection $(\Delta_Z,E_T^{\rm miss},H_T,N_j^{\max})$ & $\epsilon_{T2}$ & $\sigma_{T2}\pm\delta\sigma$ & $S$ & $B\pm\delta B$ & $S/B$ & $Z_A$ & $Z_{\rm val}$ \\ \hline
(TeV) & (fb$^{-1}$) & (GeV, GeV, GeV, --) & & (pb) & & & & & \\ \hline
8 & 20 & (10, 30, 0, 2) & $0.00295\pm0.00017$ & $0.01638\pm1.36\times10^{-5}$ & $0.9664\pm0.0562$ & $0.3923\pm0.064$ & 2.46 & 1.19 & 1.1 \\ \hline
14 & 3000 & (10, 0, 0, 1) & $0.00241\pm0.00016$ & $0.044575\pm4.49\times10^{-5}$ & $322.3\pm20.7$ & $140.2\pm24$ & 2.3 & 8.02 & 6.75 \\ \hline
27 & 15000 & (20, 30, 0, 3) & $0.00163\pm0.00013$ & $0.12135\pm1.02\times10^{-4}$ & $2967\pm2.32\times10^{2}$ & $1167\pm2.28\times10^{2}$ & 2.54 & 8.13 & 6.12 \\ \hline
100 & 20000 & (10, 50, 0, 4) & $9.00\times10^{-4}\pm9.5\times10^{-5}$ & $0.66224\pm5.86\times10^{-4}$ & $1.192\times10^{4}\pm1.26\times10^{3}$ & $5709\pm1.31\times10^{3}$ & 2.09 & 6.03 & 4.47 \\ \hline
\end{tabular}%
}
\end{table}

\begin{figure}[p]
\centering\includegraphics[width=0.92\textwidth]{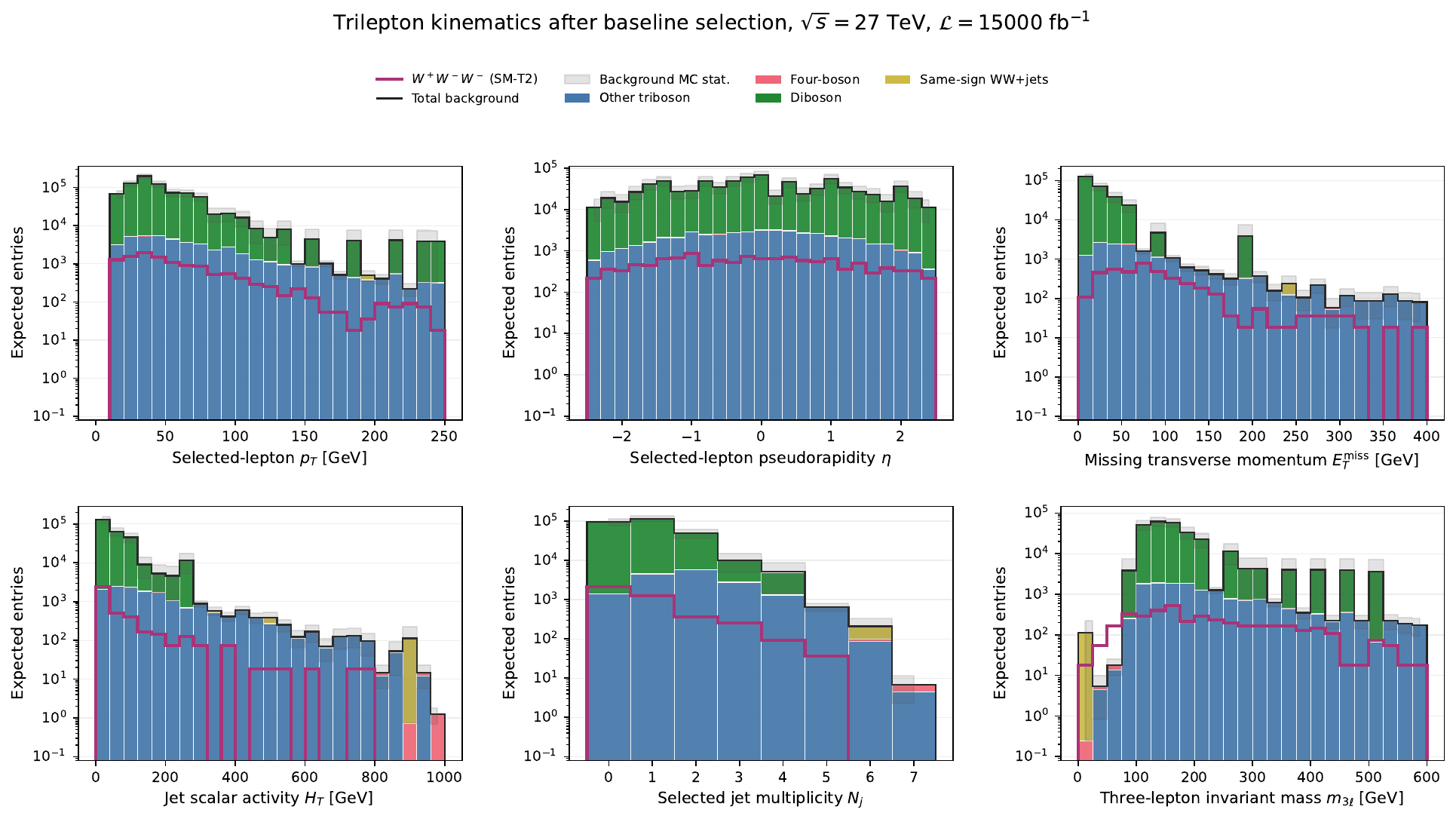}
\caption{SM-T2 baseline kinematics at 27 TeV for 15 ab$^{-1}$, before the optimized SFOS, missing-momentum, and jet requirements. The baseline requires exactly three isolated leptons, net charge $-1$, and ordered lepton thresholds 25, 15, and 10 GeV.}
\label{fig:t2kin27}
\end{figure}

\begin{figure}[p]
\centering\includegraphics[width=0.92\textwidth]{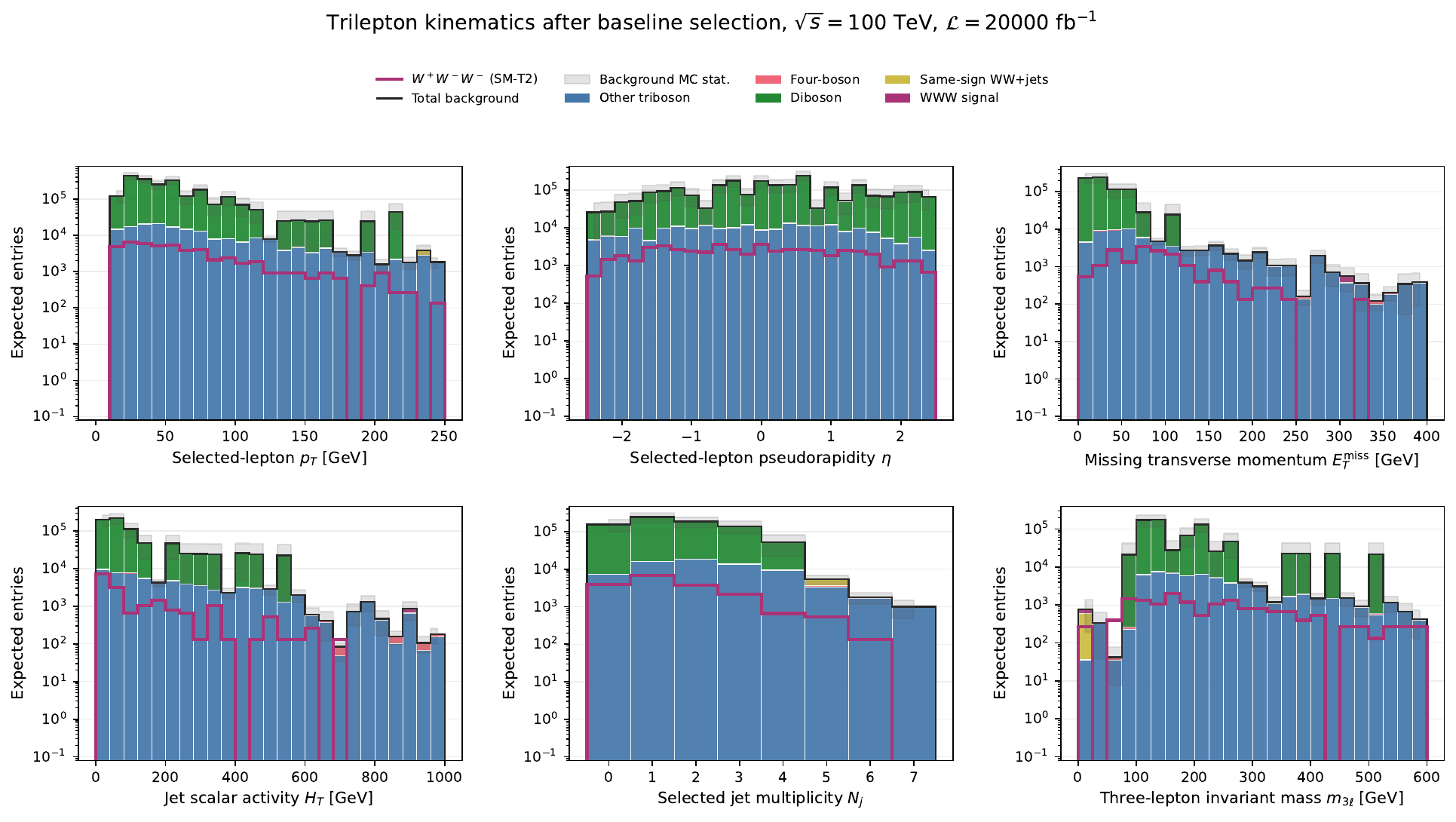}
\caption{SM-T2 baseline kinematics at 100 TeV for 20 ab$^{-1}$, using the same conventions as Fig.~\ref{fig:t2kin27}.}
\label{fig:t2kin100}
\end{figure}

Figures~\ref{fig:t2kin27} and~\ref{fig:t2kin100} show lepton transverse momentum and pseudorapidity, missing transverse momentum, jet activity, and three-lepton mass after the SM-T2 baseline selection. The preselection and plotting conventions match the SM-T1 distributions, with the net-charge requirement reversed.

\begin{figure}[ht]
\centering\includegraphics[width=0.88\textwidth]{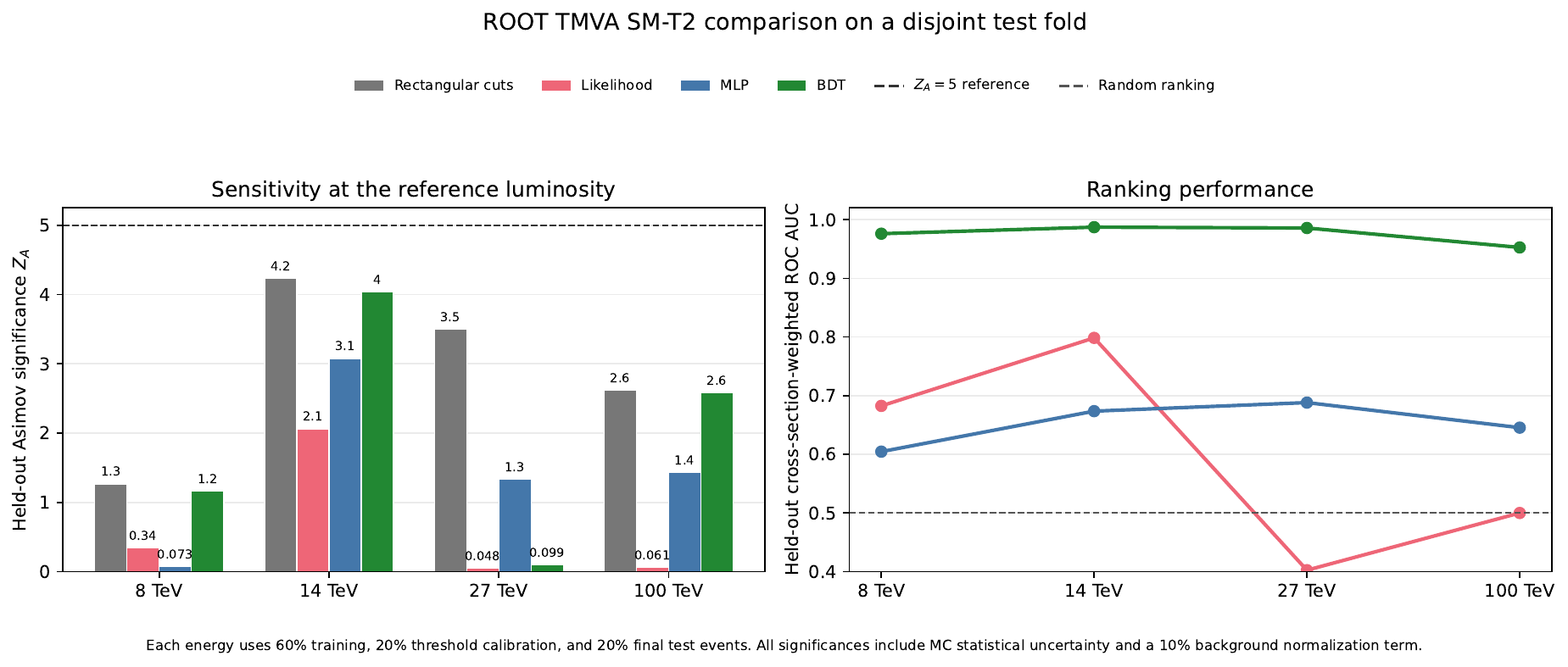}
\caption{Held-out classifier comparison for SM-T2 at the reference luminosities. The left panel gives $Z_A$ and the right panel the cross-section-weighted test ROC AUC. The $Z_A=5$ line is a visual reference; sparse test tails limit the stability of the operating-point estimates.}
\label{fig:mvat2}
\end{figure}

The same three classifiers and event-level variables were evaluated separately for SM-T2, with the charge requirement changed to $-1$. Table~\ref{tab:mvat2} and Fig.~\ref{fig:mvat2} report the held-out results. At 8 TeV the BDT nearly reaches the rectangular-cut test significance (1.16 versus 1.26), but at 14 TeV it remains below cuts (4.04 versus 4.23). At 27 TeV, despite a test AUC of 0.986, its selected test significance is only 0.10 because the chosen operating point admits a very large background yield; the MLP reaches 1.33 versus 3.50 for cuts. At 100 TeV the BDT result (2.58) is close to, but below, the cut result (2.61). Thus the high BDT AUC at 27 TeV does not translate into useful background rejection at the selected threshold. Across energies, the held-out MC counts and effective background counts are small, so these classifier comparisons remain exploratory and do not replace the T2 cut-based projections.

\begin{table}[ht]
\centering\scriptsize
\caption{Held-out TMVA comparison for the charge-conjugate SM-T2 hypothesis, using net charge $-1$. The test-fold yields and $Z_A$ are diagnostics; the sparse passing counts make the high-score operating points unstable.}
\label{tab:mvat2}
\resizebox{\textwidth}{!}{%
\begin{tabular}{|c|l|r|r|r|r|r|c|}\hline
$\sqrt{s}$ & Method & $Z_A$ & Test AUC & $S$ & $B$ & $B_{\rm eff}$ & Test MC ($n_S/n_B$) \\ \hline
(TeV) & & & & & & & \\ \hline
8 & Rectangular cuts & 1.265 & -- & 0.688 & 0.118 & 3.75 & 42/17 \\ \hline
8 & Likelihood & 0.340 & 0.683 & 0.164 & 0.175 & 6.03 & 10/43 \\ \hline
8 & MLP & 0.073 & 0.605 & 0.409 & 5.62 & 1.31 & 25/165 \\ \hline
8 & BDT & 1.163 & 0.976 & 1.11 & 0.565 & 18.38 & 68/113 \\ \hline
14 & Rectangular cuts & 4.229 & -- & 301 & 139 & 10.15 & 45/55 \\ \hline
14 & Likelihood & 2.061 & 0.798 & 93.6 & 87.5 & 7.04 & 14/44 \\ \hline
14 & MLP & 3.076 & 0.674 & 127 & 63.4 & 5.90 & 19/72 \\ \hline
14 & BDT & 4.038 & 0.987 & 281 & 146 & 11.17 & 42/71 \\ \hline
27 & Rectangular cuts & 3.500 & -- & 273 & 134 & 7.34 & 3/42 \\ \hline
27 & Likelihood & 0.048 & 0.403 & $3.19\times10^{3}$ & $2.34\times10^{5}$ & 14.18 & 35/555 \\ \hline
27 & MLP & 1.331 & 0.688 & 728 & 1.31e+03 & 8.45 & 8/75 \\ \hline
27 & BDT & 0.099 & 0.986 & $3.19\times10^{3}$ & $6.21\times10^{4}$ & 4.01 & 35/287 \\ \hline
100 & Rectangular cuts & 2.613 & -- & $7.28\times10^{3}$ & 1.69e+03 & 1.31 & 11/6 \\ \hline
100 & Likelihood & 0.061 & 0.500 & 1.92e+04 & $9.08\times10^{5}$ & 9.33 & 29/299 \\ \hline
100 & MLP & 1.432 & 0.645 & $5.30\times10^{3}$ & $6.85\times10^{3}$ & 5.39 & 8/37 \\ \hline
100 & BDT & 2.578 & 0.953 & $9.93\times10^{3}$ & $5.08\times10^{3}$ & 3.96 & 15/31 \\ \hline
\end{tabular}%
}
\end{table}

\section{Statistical treatment and luminosity dependence}
We report the signal-to-background ratio $S/B$ and the simple counting benchmarks $S/\sqrt{B}$ and $S/\sqrt{S+B}$ in the machine-readable summary. Our principal sensitivity metric is the profile-likelihood Asimov significance $Z_A$ for signal $S$, background $B$, and background variance $\sigma_B^2$:
\begin{align}
Z_A^2=2\Bigg[&(S+B)\ln\!\left(\frac{(S+B)(B+\sigma_B^2)}{B^2+(S+B)\sigma_B^2}\right)\\
&-\frac{B^2}{\sigma_B^2}\ln\!\left(1+\frac{\sigma_B^2S}{B(B+\sigma_B^2)}\right)\Bigg].
\label{eq:cowan}
\end{align}
The background variance combines a 10\% relative normalization term, the finite-MC variance, and the MG5 cross-section integration errors. The significance expression follows the Asimov profile-likelihood treatment \cite{Cowan}. The 10\% term is an analysis assumption rather than a detector-derived uncertainty. The luminosity curves in Fig.~\ref{fig:lumi} are obtained by rescaling the expected event yields with luminosity while retaining these selection efficiencies and uncertainty prescriptions.

\section{Discussion and limitations}
\label{sec:scope}
Several important backgrounds are absent from the supplied set, including top-quark production, $WZ+$jets, nonprompt or misidentified leptons, and charge misidentification. The analysis does not include pile-up, trigger efficiencies, detector-specific reconstruction corrections, or a data-driven background estimate. In particular, the zero selected events in finite diboson and $WWjj$ samples must not be interpreted as zero background rates. Their high inclusive cross sections make additional generated samples and dedicated fake/charge-misidentification estimates necessary before experimental sensitivity can be claimed.

The odd/even split provides a cut-validation check but does not replace larger independent event samples. Furthermore, the present distributions use the available Delphes reconstruction and Pythia8 boson decays; spin-correlated MadSpin decays were not used in this production chain. A full phenomenological result should regenerate incomplete and low-statistics samples, include the missing experimental backgrounds, use a detector and pile-up model appropriate to each collider, and evaluate systematic uncertainties with a dedicated likelihood model.

\section{Conclusion}
We have evaluated two charge-specific trilepton signal hypotheses using the supplied $pp\to W^+W^-W^+$ (SM-T1) and $pp\to W^+W^-W^-$ (SM-T2) samples at $\sqrt{s}=8$, 14, 27, and 100 TeV. Each selection requires exactly three isolated electrons or muons, the corresponding reconstructed net charge ($+1$ for SM-T1 and $-1$ for SM-T2), ordered lepton thresholds of 25, 15, and 10 GeV, an SFOS $Z$ veto, missing transverse momentum, and a jet-multiplicity requirement.

For SM-T1, the efficiencies are $1.86\times10^{-3}$, $7.30\times10^{-4}$, $9.40\times10^{-4}$, and $5.90\times10^{-4}$. At reference luminosities of 20 fb$^{-1}$, 3 ab$^{-1}$, 15 ab$^{-1}$, and 20 ab$^{-1}$, the expected $(S,B)$ yields are $(1.27,0.186)$, $(180,40.7)$, $(2.81\times10^3,784)$, and $(1.09\times10^4,3.79\times10^3)$, with $S/B=6.80$, 4.42, 3.59, and 2.89, and $Z_A=1.84$, 7.70, 8.62, and 6.31. For SM-T2, the efficiencies are $2.95\times10^{-3}$, $2.41\times10^{-3}$, $1.63\times10^{-3}$, and $9.00\times10^{-4}$; projected signal yields are 0.97, 322, $2.97\times10^3$, and $1.19\times10^4$, with nominal $Z_A=1.19$, 8.02, 8.13, and 6.03. Its independent validation-fold values are 1.10, 6.75, 6.12, and 4.47. The inclusive charge-channel asymmetry also falls with energy, from $A_C^{WWW}=0.3510\pm0.0005$ at 8 TeV to $0.1663\pm0.0006$ at 100 TeV, where these uncertainties include only the MG5 integration errors.

The ROOT TMVA studies for both charge hypotheses do not establish a stable improvement over rectangular cuts. For SM-T1, all classifiers are below cuts at 8 TeV; the apparent MLP gain at 14 TeV is accompanied by a test AUC of 0.570 and 15/30 passing test events; at 27 TeV the BDT gives $Z_A=3.38$, below the cut value 4.92; and at 100 TeV the BDT gives 2.42 versus 1.72 for cuts, based on 11/29 passing events. For SM-T2, the BDT values are 1.16, 4.04, 0.10, and 2.58, compared with cut values 1.26, 4.23, 3.50, and 2.61. In particular, the 27 TeV T2 BDT test AUC of 0.986 does not translate into useful background rejection at its selected threshold. Sparse held-out tails and overtraining make both MVA comparisons exploratory; cut-based estimates remain the primary projections. These conditional fast-simulation results omit important backgrounds and comprehensive detector and theory uncertainties and should not be interpreted as experimental discovery sensitivities.

\begin{acknowledgments}
The authors acknowledge the use of MadGraph5\_aMC@NLO, Pythia8, Delphes, ROOT, and FastJet in the event-production and analysis workflow.
\end{acknowledgments}

\end{document}